\documentclass[
    ,final            
  ]
  {aipproc}

\layoutstyle{6x9}
\usepackage{amssymb}
\usepackage{subfigure}

\begin{document}

\title{Self-gravitating discs with radiative transfer: their role in giant planet formation}


\author{Farzana Meru}{
  address={Astrophysics Group, School of Physics, University of Exeter, Stocker Road, Exeter, EX4 4QL}
}

\author{Matthew R. Bate}{
  address={Astrophysics Group, School of Physics, University of Exeter, Stocker Road, Exeter, EX4 4QL}
}

\begin{abstract}
We present preliminary results on the ability of self-gravitating discs to cool in response to their internal heating.  These discs are modelled using a Smoothed Particle Hydrodynamics (SPH) code with radiative transfer \cite{WH_Bate_Monaghan2005} and we investigate the ability of these discs to maintain a state of thermal equilibrium with their boundaries as an indication of their likelihood to fragment.

\end{abstract}

\maketitle


\section{Introduction}
Gravitational instability has been proposed as a theory for giant planet formation \cite{GI_Cameron, Boss_GI}, involving massive discs in the early stages of their evolution, and is particularly favoured for its rapid formation timescales.  The stability of a self-gravitating disc can be described by the stability parameter \cite{Toomre_stability1964}, $Q=\frac{c_s\kappa}{\pi\Sigma G}$, where $c_s$ is the sound speed in the disc, $\kappa$ is the epicyclic frequency, which for Keplerian discs is approximately equal to the angular frequency, $\Omega$, $\Sigma$ is the surface mass density and $G$ is the gravitational constant, and where a value $\lesssim 1$ implies instability.  Once the surface mass density and the rotation of the disc have been established, the stability is purely dependent on the disc temperature.  \citet{Gammie_betacool} showed that in addition to this, a fast cooling rate is also required in order to overcome the stabilising compressional heating.

The early evolution of massive self-gravitating discs has been considered using simplified cooling parameters to describe the thermodynamics \cite[e.g.][]{Lodato_Rice_original} and using grid-based and hydrodynamical radiative transfer calculations \cite[e.g.][]{Boss_GI_RT, Cai_etal_GI_RT, Mayer_etal_GI_RT} though results for the latter have differed.  We use SPH with flux limited diffusion \cite{WH_Bate_Monaghan2005} to model the evolution of such radiative transfer discs to determine the conditions more likely for fragmentation.

\section{Disc Setup}
We model a $25 \rm{AU}$, $0.1 \rm{M_\odot}$ disc using 250,000 SPH particles, surrounding a $1\rm{M_\odot}$ star, modelled using a sink particle.  We use temperature and surface mass density profiles of $T\propto R^{-\frac{1}{2}}$ and $\Sigma\propto R^{-1}$ respectively, such that the aspect ratio is initially $\sim 0.05$.  The temperature of the disc's vertical boundary reflects the heating from the star and this boundary is applied for optical depths of $\tau \lesssim 1$.  It is also kept at a constant temperature and is initially in a state of thermal equilibrium with the midplane.

\section{Results \& Discussion}

\begin{figure}
  \begin{minipage}[b]{0.5\linewidth} 
    \centering
    \includegraphics[height=5.28cm]{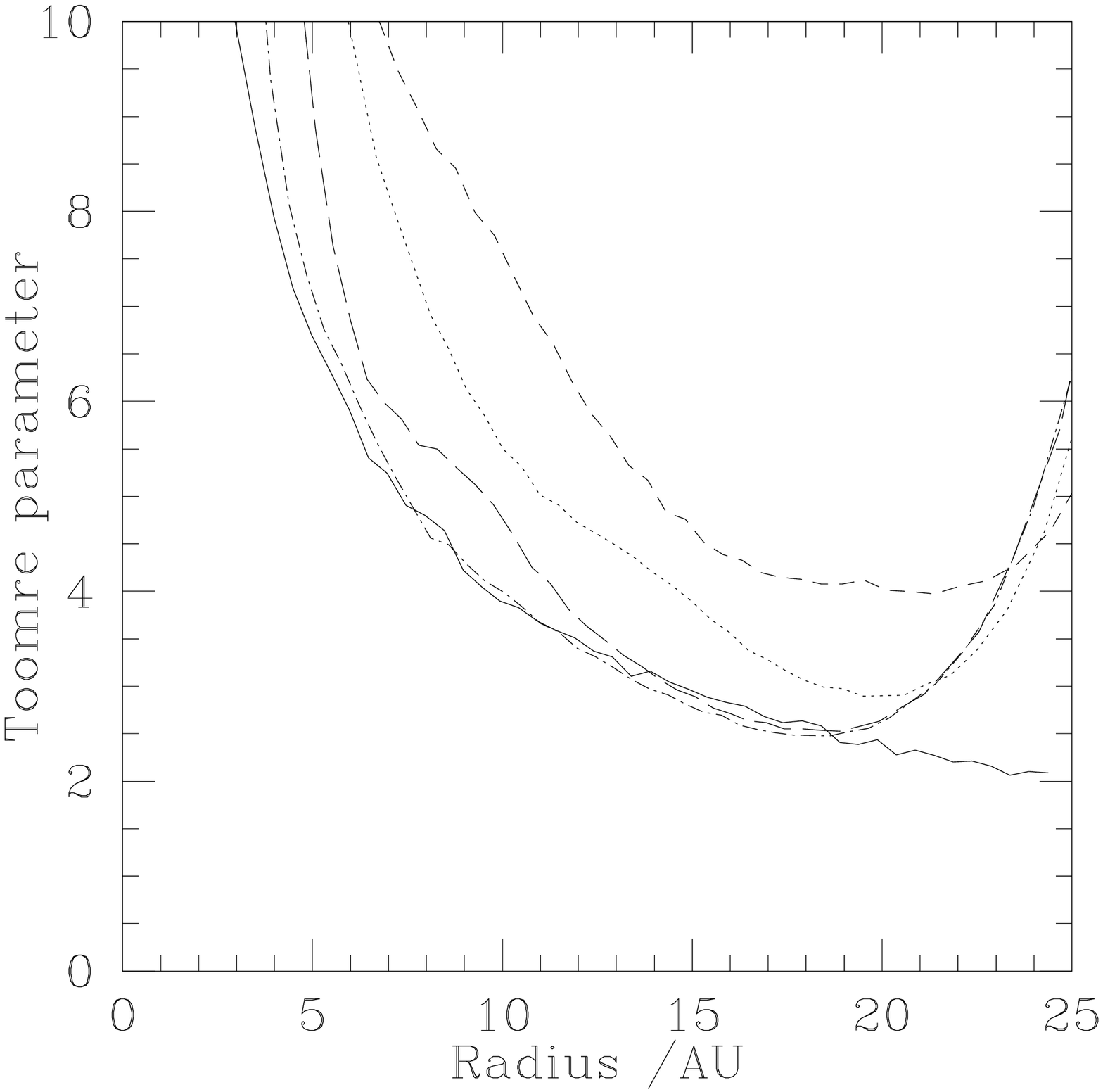}
  \end{minipage}
  \hspace{0.1cm} 
  \begin{minipage}[b]{0.5\linewidth}
    \centering
    \includegraphics[width=5.28cm]{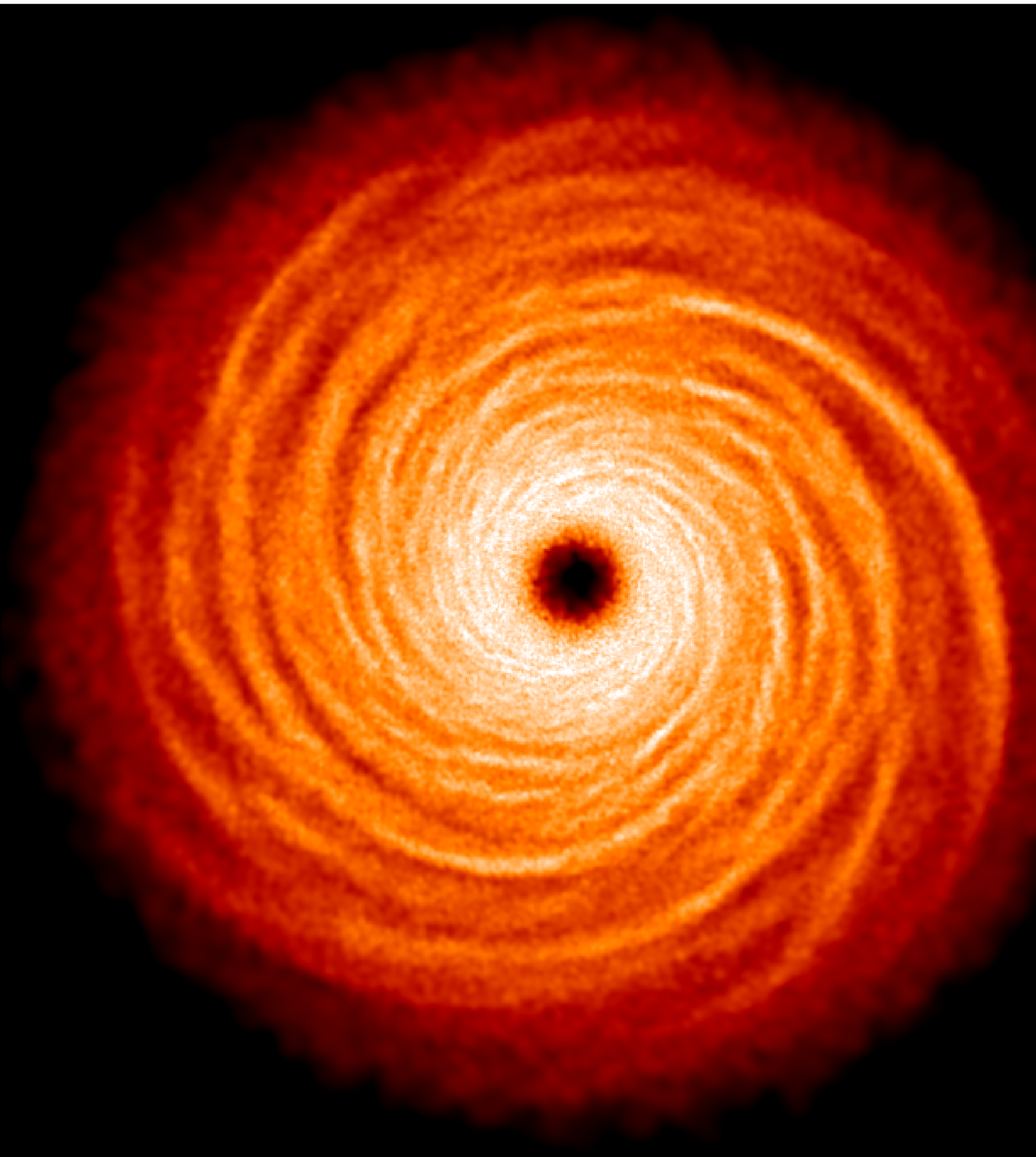}
    \caption{\textbf{a.}~Azimuthally averaged Toomre values for the initial discs (solid) and at $t=6.4 \rm{ORP}$ for $10x$ (short-dashed), $1x$ (reference; dotted), $0.1x$ (long-dashed) \& $0.01x$ (dot-dashed) interstellar opacities.  \textbf{b.}~Simulation image of a marginally stable self-gravitating disc showing high (white) to low (red) density.}
    \label{figure}
  \end{minipage}
\end{figure}

The key result is that modelling discs more realistically with radiatve transfer reduces the likelihood of fragmentation in self-gravitating discs.  Given that the boundary temperature profiles have been set, if the discs are able to respond rapidly enough to the internal heating, the Toomre profiles at the end of the simulations will be the same as the initial Toomre profiles.  If the discs heat up and cannot even maintain thermal equilibrium with their vertical boundaries (and heat up), fragmentation is not likely.

Figure~\ref{figure}a shows that the reference and high opacity discs heat up due to internal processes and cannot cool fast enough to regain thermal equilibrium with their boundaries.  However, the very low opacity disc is able to cool rapidly enough in order to maintain a state of thermal equilibrium with the disc boundary.  This is also the case with a larger ($300 \rm{AU}$), cooler disc, suggesting that given more favourable (cooler) boundary conditions, these discs may potentially cool quickly enough to fragment.  In fact, we find that, indeed, combining these properties, a larger, cooler, low opacity disc is able to fragment.

We are currently investigating the effect of reducing the boundary temperature during the simulations in order to determine whether it is realistically possible to obtain marginally stable discs such as that shown in figure~\ref{figure}b, or to drive them to fragment.

\bibliographystyle{aipproc}   

\bibliography{meru_bibtex}

\IfFileExists{\jobname.bbl}{}
 {\typeout{}
  \typeout{******************************************}
  \typeout{** Please run "bibtex \jobname" to optain}
  \typeout{** the bibliography and then re-run LaTeX}
  \typeout{** twice to fix the references!}
  \typeout{******************************************}
  \typeout{}
 }

\end{document}